\documentclass[sigconf]{acmart}

\AtBeginDocument{%
  \providecommand\BibTeX{{%
    \normalfont B\kern-0.5em{\scshape i\kern-0.25em b}\kern-0.8em\TeX}}}

\copyrightyear{2024} 
\acmYear{2024} 
\setcopyright{acmlicensed}\acmConference[WCCCE '24]{The 26th Western Canadian Conference on Computing Education}{May 2--3, 2024}{Kelowna, BC, Canada}
\acmBooktitle{The 26th Western Canadian Conference on Computing Education (WCCCE '24), May 2--3, 2024, Kelowna, BC, Canada}
\acmDOI{10.1145/3660650.3660659}
\acmISBN{979-8-4007-0997-5/24/05}

\usepackage{multirow}
\usepackage{makecell}
\usepackage{xcolor}

\definecolor{ggred}{HTML}{F8766D}
\definecolor{ggblue}{HTML}{00bfc4}

\usepackage{booktabs}
\begin{document}


\title{Opportunities for Adaptive Experiments to Enable Continuous Improvement in Computer Science Education}

\author{Ilya Musabirov} 
\affiliation{
\institution{University of Toronto}
  \city{Toronto}
  \country{Canada}
}
\email{imusabirov@cs.toronto.edu}

\author{Angela Zavaleta Bernuy} 
\affiliation{
\institution{University of Toronto}
  \city{Toronto}
  \country{Canada}
}
\email{angelazb@cs.toronto.edu}

\author{Pan Chen} 
\affiliation{
\institution{University of Toronto}
  \city{Toronto}
  \country{Canada}
}
\email{pan.chen@utoronto.ca}

\author{Michael Liut} 
\affiliation{
\institution{University of Toronto Mississauga}
  \city{Mississauga}
  \country{Canada}
}
\email{michael.liut@utoronto.ca}

\author{Joseph Jay Williams} 
\affiliation{
\institution{University of Toronto}
  \city{Toronto}
  \country{Canada}
}
\email{williams@cs.toronto.edu}



\begin{abstract}
Randomized A/B comparisons of alternative pedagogical strategies or other course improvements could provide useful empirical evidence for instructor decision-making. However, traditional experiments do not provide a straightforward pathway to rapidly utilize data, increasing the chances that students in an experiment experience the best conditions. Drawing inspiration from the use of machine learning and experimentation in product development at leading technology companies, we explore how adaptive experimentation might aid continuous course improvement. In adaptive experiments, data is analyzed and utilized as different conditions are deployed to students. This can be achieved using machine learning algorithms to identify which actions are more beneficial in improving students' learning experiences and outcomes. These algorithms can then dynamically deploy the most effective conditions in subsequent interactions with students, resulting in better support for students' needs. We illustrate this approach with a case study that provides a side-by-side comparison of traditional and adaptive experiments on adding self-explanation prompts in online homework problems in a CS1 course. This work paves the way for exploring the importance of adaptive experiments in bridging research and practice to achieve continuous improvement in educational settings.

\end{abstract}




\maketitle

\section{Introduction}

Instructors and designers of digital learning experiences typically make numerous decisions about which pedagogical strategies to use, often drawing on their anecdotal experiences and judgment. For example, students may be presented with hundreds of programming practice problems in a given course. Instructors may want to know if asking students to write explanations for their answers after solving some of these problems is worthwhile. Randomized A/B comparisons can be used within the context of a course to compare alternative pedagogical strategies on specific problems by collecting data from the actual student population. Increasing efforts have been made to provide tools and strategies for instructors and researchers to carry out such experiments \cite{reza_mooclet_2021, ritter_third_2022, heffernan_assistments_2014}.

However, for traditional experiments, there is no clear pathway to use data from the experiment to rapidly improve the educational experiences of students in the course. In some cases, there are valid concerns that randomized experiments may be unfair to students. Moreover, the vast majority of experiments are conducted to be published in research venues, and may never have a real impact on the educational settings in which they were conducted in the first place. In contrast, in industry, there is a growing body of best practices for adapting the principles of experimental research and optimization not only to produce knowledge but also to generate systematic continuous improvement of real-world products \cite{agarwal_making_2016}. A widely used approach is to apply machine learning algorithms to analyze data from experiments in real-time and dynamically \textit{adapt} the experiment to provide more effective conditions, which are also called ``arms,'' to future participants, with a probability changing as there is greater certainty about what leads to better outcomes.

In this paper, we discuss the adaptive experiments approach, and its potential benefits for course improvement using a case study of an adaptive experiment in a CS1 course. There are two main goals. We compare data from adaptive and traditional experiments, \textbf{\underline{G1}: Demonstrating how conducting an adaptive vs. traditional experiment in a CS1 course can benefit students}. We also \textbf{\underline{G2}: Discuss some promising areas for applying adaptive experiments to continuous improvement in CS courses}.

Our case study intervention was conducted in the Fall 2021 semester in the Introduction to Computer Programming course (CS1) at a public research-focused North American university. Students with varying STEM backgrounds from different departments were able to enroll in the course. However, most of the students were first-years who wanted to pursue a computer science degree. With more than 1,000 students initially enrolled in the course, the course provides a sample sufficient to confidently test different pedagogical approaches with large to medium effect sizes. However, splitting the student population 50\%-50\% between different experimental conditions increases the potential ethical impact in the case where one of the conditions is substantially better than the other. The goal of an adaptive experiment is to reach a trade-off between collecting the data needed to verify the hypotheses about the best learning experience, and benefiting students in the current course. This experiment was approved by the institution's local ethics board.

The paper makes two contributions. Firstly, this work presents a real-world proof-of-concept deployment of an experimental paradigm, namely testing prompts on online homework problems. As such, this work reflects on how adaptive experimentation based on the Bayesian Multi-Armed Bandit algorithm Thompson Sampling~\cite{agrawal_analysis_2012}, could be executed and altered. Secondly, this work compares traditional and adaptive experiment approaches. We provide researchers and instructors with an example of an algorithm deployed in one context and reflect on its uses, benefits, and limitations. We hope that the paper can be the first step towards a body of real-world adaptive experimentation case studies that will bridge research and practice for continuous improvement in computer science classes.

This paper continues by outlining related work on continuous improvement, design science, and action research in education, as well as on A/B comparisons in the technology industry and the challenges of their use in education, in Section~\ref{sec:related}. Section~\ref{sec:ts} illustrates how Multi-Armed Bandits work, using the example of Beta-Bernoulli Thompson Sampling. Section~\ref{sec:cs-SelfExpl} then presents a case study focusing on adaptive experiment design, followed by data analysis, results, and a comparison with a traditional experiment. Finally, Section~\ref{sec:opportunities} discusses the opportunities and limitations surrounding A/B comparisons, and Section~\ref{sec:conclusion} summarizes the research and describes future directions in this space.

\section{Related Work}
\label{sec:related}

\paragraph{Continuous Improvement, Design Science, and Action Research in Education}

Although the problem of effective Instructor-Researcher collaborations in computer science education research may seem trivial at first glance, real-world partnerships require reformulating models of traditional research \cite{penuel_conceptualizing_2015}. The gap between computer science (CS) education research and the adoption of its results is wide despite the increase in the proliferation of evidence-informed education practices \cite{hovey_survey_2019}. Although some subfields of educational research, such as Learning Analytics, are aimed at supporting decision-making in practice \cite{du_systematic_2021,hilliger_design_2020,mattingly_learning_2012}, they rarely use experimental evidence to support continuous improvement processes. Even when instructors and researchers have the opportunity to work together, there is a clash between traditional linear research logic and real-world problems. In these environments, new ideas or context changes can emerge in the process and need to be addressed rapidly \cite{hallinan_getting_2011}. Often, the gap between research results and their implementation into practice is too wide, and instructors are overwhelmed by current work, lacking time and resources to implement changes \cite{hovey_survey_2019,barker_what_2015}. 

There are many promising ideas and approaches to overcome these challenges, including design research \cite{martin_intelligent_2022,mckenney_conducting_2018} and action research in education \cite{clement_call_2004}, network improvement communities \cite{hilliger_design_2020}, and research-practice partnerships \cite{penuel_conceptualizing_2015}. However, they can arguably all benefit from technologies that support not only approaches to collect evidence but also the rapid implementation of uncovered solutions into practice~\cite{marwan_adaptive_2020}.

\paragraph{Systematic A/B Comparisons in Technology Industry and Other Fields}

In the technology industry, specifically in product development, systematic A/B testing has become the standard way of discovery and improvement, bringing researchers and practitioners together to find best practices \cite{kohavi_trustworthy_2020} and build infrastructures \cite{deng_b_2017} aimed at supporting the continuous improvement of products and services. In addition to investing in custom experimental software infrastructure, many third-party tools and services allow the integration of experimental capabilities on the web, mobile, and other platforms. Successful cases also demonstrate the benefits of systematic A/B testing in a wide range of areas, from public policy \cite{azevedo_b_2018} to medicine \cite{kohavi_online_2020, berliner_senderey_its_2020}. Companies such as Meta \cite{bakshy_ae_2018} and Microsoft \cite{agarwal_making_2016} pioneer the application of the adaptive approach, developing and exploiting machine learning-based infrastructures. These infrastructures support thousands of experiments in their products and services.

\paragraph{Challenges of Systematic A/B Comparisons in Education}

In education, the situation is quite different. A recent review of empiricism in CS education research \cite{heckman_systematic_2022} states that more than 80\% of the articles in the main venues of CS education research report at least some empirical evaluation; however, the share of works using experimental methods is much lower \cite{lishinski_methodological_2016}. The picture with a systematic approach to A/B comparisons to support continuous improvement at scale differs even more from industry and policy.

Besides general barriers and costs related to experimental research, access to infrastructure is one of the key factors for adopting systematic A/B comparisons. Mainstream Learning Management Systems and online course platforms do not yet provide default instructor-directed infrastructures for A/B comparisons \cite{williams_enhancing_2018}. There are also limitations to adopting traditional web experimentation platforms \cite{williams_enhancing_2018}, including privacy and ethical considerations. Therefore, the growing efforts of the educational community and science policymakers in the last decade have been directed toward creating infrastructures, gathering best practices, and disseminating knowledge through forums for researchers and practitioners \cite{stamper_rise_2012,ritter_third_2022}. 

Multiple recent funding efforts and competitions, such as the XPRIZE Digital Learning Challenge\footnote{\url{https://www.xprize.org/challenge/digitallearning}}, the Learning Engineering Tools Competition\footnote{\url{https://tools-competition.org}}, and NSF grants for creating research infrastructures, had aimed to close the gap in education. A growing number of educational experimentation platforms have become available to instructors and researchers, for example, UpGrade \cite{fancsali_closing_2022}, MOOClet \cite{reza_mooclet_2021}, E-TRIALS \cite{heffernan_assistments_2014}, and Terracotta \cite{motz_embedding_2018}, some of them building adaptive experimentation capabilities.

There are some promising examples of adaptive experiments that rely on these new infrastructures to perform systematic A/B comparisons. In \cite{zavaleta-bernuy_using_2021} and \cite{yanez_increasing_2022}, researchers report on the use of adaptive experiments to improve engagement with course emails. \cite{asano_exploring_2021} explores ways to support students in an online homework system adaptively. 
As such, this work aims to expand the range of such studies by outlining the emerging principles, benefits, and challenges of adaptive experimentation in computer science classrooms.

\section{Adaptive Experiments Using Thompson Sampling}
\label{sec:ts}

One of the key approaches to Adaptive Experimentation is based on Multi-Armed Bandit (MAB) algorithms.

In this section, we describe the Thompson Sampling MAB algorithm used in our case study. This algorithm was shown to have beneficial theoretical properties, particularly regret bounds, and good performance in practical problems \cite{chapelle_empirical_2011,agrawal_analysis_2012}. Moreover, it is possible to ensure the statistical interpretability of the algorithm at different steps, as we will illustrate in this section.

To make the paper more useful as a starting guide for instructors and researchers, we also demonstrate the work of the algorithm first with artificial examples and then, in the next Section, as part of the case study in an authentic classroom setting.

\emph{Thompson Sampling} learns how effective each of the conditions is, e.g., showing or skipping self-explanation prompts. This online learning happens based on the feedback (\emph{reward} in MAB terms, e.g., correct or incorrect answer on the following task) and happens from every interaction of participants with the algorithm without the need to gather all evidence in advance. When the reward is binary, the Beta Bernoulli version \cite{chapelle_empirical_2011} of the algorithm (TS-BB) can be used. 

\begin{figure}[ht]
    \centering
    \includegraphics[width=220px]{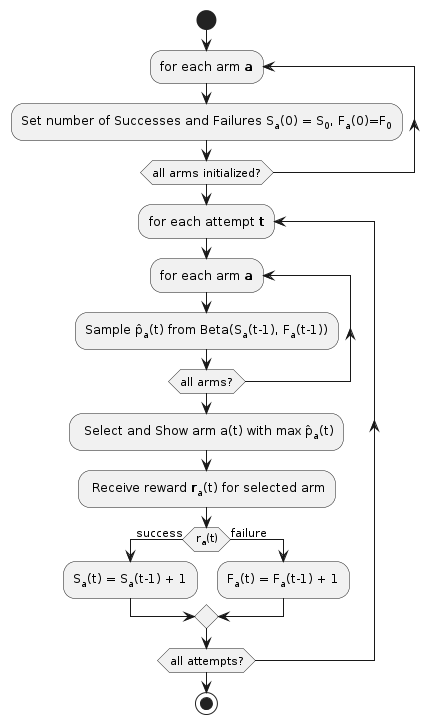}
    \caption{Schematic implementation of Thompson Sampling Beta Bernoulli algorithm (TS-BB), based on \cite{chapelle_empirical_2011}.}
    \label{fig:tsbb}
    \Description{Schematic implementation of Thompson Sampling Beta Bernoulli algorithm (TS-BB), based on \cite{chapelle_empirical_2011}.}
\end{figure}

The main steps of the schematic implementation of TS-BB are illustrated in Figure~\ref{fig:tsbb}.

The probability distributions, representing our knowledge about each arm (condition), are initialized with priors, which reflect our initial beliefs. For example, we can set them with respect to the numbers of previously observed \textbf{S}uccesses and \textbf{F}ailures. If no prior information is assumed, the prior can be set to correspond to $\textbf{S}_0 = 1$ and $\textbf{F}_0 = 1$, reflecting equal chances of success or failure and high uncertainty (Uniform Random distribution).

Then, for each attempt and for each arm, the algorithm samples a value from each corresponding probability distribution\footnote{Beta distribution $p_a(t) \sim \text{Beta}(S_a(t-1), F_a(t-1))$ (conjugate to Bernoulli), with $a$ denoting the arm and $t$ indexing the attempt}. These values are used to find the arm, corresponding to the maximal expected reward, and show it to the participant. Lastly, the algorithm records the reward for this arm and uses it to update the corresponding probability distribution (updating the number of \textbf{S}uccesses or \textbf{F}ailures).

\paragraph{Burn-in period and batch size}

Additional parameters that allow better balancing of statistical inference and optimization properties are the burn-in period and the batch size. The burn-in period at the beginning of the study sets the adaptive experimentation engine to assign the first $k$ cases (or batches) to be served using uniform random assignment instead of Thompson Sampling, while still using these cases to calculate rewards and update distributions. This allows us to achieve less biased samples if we plan to apply null hypothesis significance testing to our data.
The batch size regulates the minimal number of data points collected by the adaptive experimentation engine before updating the parameters, thereby reducing potential noise and computational effort.

Note that these two additional parameters, the burn-in period and the batch size, are not only useful to balance statistical inference and continuous improvement interests but also can contribute to alleviating an early- vs. late-comer problem. Thompson Sampling, as well as many online learning and dynamic experimental design approaches, behaves well under stationarity, i.e., the main properties of our distribution do not change over time. However, sometimes, we can make a reasonable assumption that students interacting with our system early on are somewhat different in behaviour from those starting interaction later. For example, early homework solvers may differ from late-comers. Setting up a reasonable burn-in period can partially alleviate this issue by preventing premature swings to a condition preferred by early-comers until reasonable evidence is gathered.
\paragraph{Illustrative examples} Next, we will illustrate the potential behaviour of the algorithm in two examples of in-progress adaptive experiments with two conditions (Table~\ref{tab-examples}). For each of the examples, let us assume that we have a history based on 20 interactions with students.

\begin{table}[htb]
   \centering
\begin{tabular}{@{}cccc@{\hspace{5pt}}l@{}}
  \toprule
  \makecell{Condition\\(arm)} & Successes & Failures & EV($p_a(t)$) & Distributions \\ \midrule
  \multicolumn{5}{c}{Example 1} \\ 
   \midrule
  \colorbox{ggred}{1}  & 4 & 6 & 0.417 & \multirow{2}{*}{\includegraphics[width=60px]{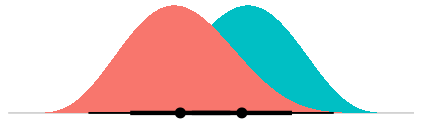}} \\
  \colorbox{ggblue}{2} & 6 & 4 & 0.583 &  \\ 
  \midrule 
  \multicolumn{5}{c}{Example 2} \\ 
  \midrule
  \colorbox{ggred}{1} & 2 & 8 & 0.25 & \multirow{2}{*}{\includegraphics[width=60px]{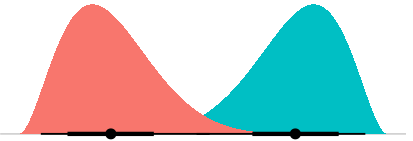}} \\
  \colorbox{ggblue}{2} & 8 & 2 & 0.75 &  \\
  \bottomrule
\end{tabular}
\caption{Two illustrative examples of in-progress adaptive experiments with (1, 1) priors. The Distributions column shows the sketches of the probability distribution for the two conditions (arms) in each example.}
\label{tab-examples}
\end{table} 

In both examples, the data on the successes and failures for each condition favours \colorbox{ggblue}{Condition 2} over \colorbox{ggred}{Condition 1} (Table~\ref{tab-examples}). However, in Example 2, we have more compelling evidence to select Condition 2. We can build intuition about the differences between the examples by looking at small-multiple visualizations in the Distributions column: in Example 2, the distributions for each condition are more sharply peaked than in Example 1, indicating lower uncertainty, and the difference between their centers (expected values of each condition) is greater.

For TS-BB, during the next attempt in both examples, we are more likely to have $\hat{p_2} > \hat{p_1}$ and show Condition 2 to students. However, in Example 1, TS-BB is likely to explore more as our sampling can sometimes still lead to selecting Condition 1. With each additional interaction, our estimates for arm distributions will become more precise. Assuming that the observed pattern continues and as Condition 2 demonstrates better results, we are likely to serve it more frequently in new interactions.

\paragraph{Selecting Reward} Another key component of adaptive experiment design with Thompson Sampling is the choice of reward. For adaptive experiments, a good choice of reward is one of the intervention's proximal outcomes, typically observed right after the experiment, e.g., the correctness of the student's response to the follow-up task. This implicitly sets higher requirements for the quality of assessment elements: we want the tasks used as rewards to have sufficient quality to differentiate students' abilities and to be strongly correlated with course outcomes.
 
In summary, the benefit of using proximal outcomes as algorithm rewards is the increased speed of learning and continuous improvement. It allows the algorithm to learn at every step, increasing the probability of assigning the arm expected to produce the highest reward. This leads to one important opportunity, in contrast to traditional experiments: students can benefit from the current findings as we continue to collect data, update, and change what students receive with every interaction, balancing exploration and exploitation. Thus, we automate quick decisions, such as ``what to show students'', without requiring the input of the instructor or researcher. This allows students to have a better educational experience when they actually require it: as they are learning and in the mindset of solving specific course problems. Early attempts provide data that helps future students or even themselves if they are to solve the same problem again.

\section{Case Study: Integrating Self-Explanations}
\label{sec:cs-SelfExpl}
We analyze a case study of an adaptive experiment in which instructors and researchers collaborate. In this situation, they aim to explore whether adding a self-explanation prompt improves performance on a CS1 homework problem, while also immediately using data from students to provide the current best condition.

\paragraph{Technical Setup for Algorithm Comparison}

One of the goals of this paper is to demonstrate side-by-side adaptive vs. traditional experimental approaches. To do that, we employed additional features of our open-source adaptive experimentation framework \cite{reza_mooclet_2021}, allowing us to use two different policies for the assignment of participants in parallel. For each new participant, the framework chooses either a traditional experimental assignment policy (based on Uniform Random (UR)) or an adaptive experimental assignment policy (based on the Thompson Sampling Beta Bernoulli Algorithm (TS-BB)). The probability of being assigned to the UR or TS-BB policies is 0.5 for each participant. For UR (a traditional experimental assignment policy), the framework randomly selects one of the two conditions previously described. In the case of TS-BB (an adaptive experiment policy), the choice is guided by the TS-BB algorithm.
This setup allows us to demonstrate the similarities and differences between the algorithms, enabling statistical and machine learning researchers to study their properties. However, a classroom deployment can exclusively use an adaptive approach.

\paragraph{Case Study Setting}

\begin{figure}[ht]
    \centering
    \includegraphics[width=230px]{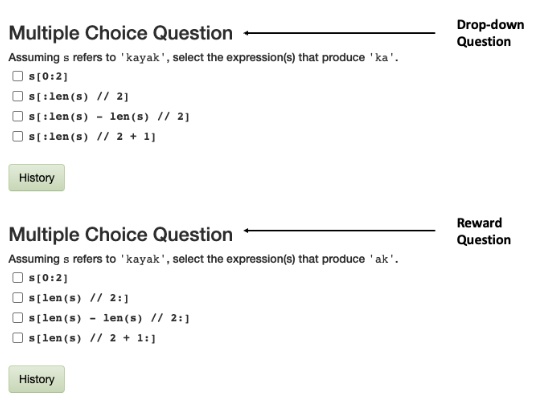}
    \caption{Multiple-choice Questions connected to the drop-down in the online homework system.}
    \Description{Multiple-choice Questions connected to the drop-down in the online homework system.}
    \label{fig:mcqs}
\end{figure}

\begin{figure}[htb]
    \centering
    \includegraphics[width=230px]{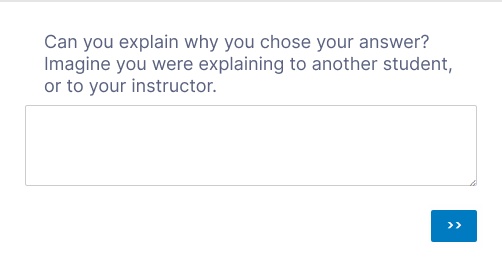}
    \caption{Design of the self-explanation prompt drop-down.}
    \Description{Design of the self-explanation prompt drop-down.}
    \label{fig:drop}
\end{figure}

As part of a CS1 course, students were required to complete weekly homework problems to keep up with the course content. Our setup involves integrating our A/B testing framework within the existing online homework system. We display the experimental condition as a drop-down~(e.g., Figure \ref{fig:drop}), which appears after a student submits a response to one of the multiple-choice questions (MCQs, e.g., Figure~\ref{fig:mcqs}, top). We then present a second MCQ (e.g., Figure~\ref{fig:mcqs}, bottom), which measures the same learning objective, immediately after the student completes the drop-down question, providing the reward: the correctness of the student's answer.

\paragraph{Experimental Design}
The experiment contains two conditions: Condition 1, which does not show a self-explanation prompt, and Condition 2, which shows the Self-Explanation prompt (Figure \ref{fig:drop}) -- \textit{Can you explain why you chose your answer? Imagine that you were explaining to another student, or to your instructor.} The drop-down is shown immediately after a student submits their answer to the MCQ (Figure \ref{fig:mcqs}, top) and appears to students regardless of whether their answer is correct or incorrect. After finishing the first MCQ and submitting the answer to the drop-down prompt, they proceed to the second MCQ (Figure \ref{fig:mcqs}, bottom). The correctness of the student's response to this MCQ is then used as a reward for the algorithm.

\paragraph{TS-BB settings}

In this scenario, the burn-in period was set at 100 to achieve better statistical properties of the data, indicating that the first 100 students were randomly assigned uniformly to one of the conditions. The batch size was set to 10, meaning that the algorithm updated the parameters (including assignment probabilities) every 10 interactions with the experiment.
%

\paragraph{Analysis and Results}

In our case, 483 students (52\%) were assigned to the traditional experiment, and 438 (48\%) were assigned to the adaptive experiment.
For the traditional experiment, there is a significant difference between the means with Condition 2 (self-explanation prompt) $M_2 = 0.608 (SEM=0.032)$ and Condition 1 (no prompt) $M_1 = 0.512 (SEM=0.032)$. Wald Z statistic = $2.122$. As a result, this provides some evidence that the self-explanation prompt, in fact, influences the accuracy of a subsequent problem, with typical caveats: the Type I error rate of $5\%$ emphasizes the chance that such data could be generated even if there were no differences between conditions. Moreover, such an effect might not replicate because of changes in context, such as a new class, or other changes to the course. Together with the instructors, we could decide at the end of the experiment whether to include a self-explanation prompt as a default option or to plan to test subsequent hypotheses in our line of inquiry.

In the adaptive experiment, the computed estimates were $M_2 = 0.599$ and $M_1 = 0.539$, also favouring self-explanation prompts. However, the main benefit of adaptive experimentation in this setting is the difference in \emph{allocation of participants to different conditions}. 
In the traditional experiment, as expected, we have almost equal assignments to arms ($N_1 = 246$, $N_2 = 237$). In contrast, in the adaptive experiment, the assignment proportions are quite different. Only $76$ students ($17.4\%$) were assigned to Condition 1 (the worse condition). That means that the algorithm served $262$ ($82.6\%$) students with a self-explanation prompt (Condition 2, the best option according to our analysis). This corresponds to more than a 30\% reallocation toward a better outcome, illustrating how adaptive A/B comparisons can support the rapid use of gathered evidence, which is especially relevant for real classrooms.

\section{Discussion: Opportunities and Limitations of Adaptive Experimentation}
\label{sec:opportunities}

\textbf{\underline{G1}: How conducting an adaptive vs. traditional experiment in a CS1 course can benefit students}
This paper presents a comparison of a real-world adaptive deployment and a traditional experiment. We demonstrate that one paradigm is promising, in which a range of adaptive experiments could be conducted for continuous improvement: different kinds of drop-down messages, prompts, and supports, while students are doing online practice problems.

We also outline a specific Bayesian Multi-Armed Bandit algorithm, Thompson Sampling, which aims to assign better conditions based on available data. It provides an intuitive and interpretable approach to adapting an experiment, allowing instructors to make decisions based on ongoing experimental results or rely on the algorithm to continuously improve student experiences. We showed how an adaptive algorithm can put as many as $82.6\%$ of the students in a better condition, improving from $50\%$ for the traditional assignment. This can enable a continuous improvement approach analogous to that used in product management by software companies to enhance the product experience \cite{bakshy_ae_2018, agarwal_making_2016}.

Although applications of machine learning algorithms, such as Thompson Sampling, in experiments afford new features for implementing a continuous course improvement process, they introduce new considerations related to statistical inference.

As the randomization of assignment to different arms is not independent, applying traditional statistical tests to TS data can lead to an increase in the False Positive Rate and/or a decrease in statistical power \cite{williams_challenges_2021,rafferty_statistical_2019}. There are two primary directions for active research in this area by statisticians and methodologists. The first focuses on analyzing and developing statistical tests better suited to bandit data, such as Adaptively Weighted Inverse Probability Weighting (AW-IPW) \cite{hadad_confidence_2021} and the Allocation Probability test \cite{deliu_efficient_2021}. The second focuses on extending machine learning algorithms to produce data with better statistical properties. One of the simpler approaches we mentioned earlier is adding a burn-in period. Others include algorithms like TS-PostDiff \cite{li2022algorithms}, which blend traditional experiments with adaptive experiments, using Bayesian statistics and machine learning.

This paper did not cover another crucial opportunity for the modern classroom that is easy to implement with adaptive experimentation—personalization. Using an extension of TS—Contextual TS, we can personalize the improvement process for different groups of students or situations, automatically discovering what works best for whom.

\textbf{\underline{G2}: Promising areas for conducting adaptive experiments in CS1}
Although adaptive experimentation, like traditional experiments, can be used in many venues of CER, there are some natural areas where rapid continuous improvement through adaptive experimentation can be especially beneficial.
Instructors and researchers can imagine and evaluate many types of support based on what they might provide in person, such as hints that vary in the amount of information, links to resources that cover key concepts, discussion forums where students can ask questions, and embedded questions that students have previously asked.

From a content perspective, one promising application area is the rapid evaluation of alternative explanations for problems and approaches to support problem-solving~\cite{leinonen_exploring_2021,williams_axis_2016}. The rapidness and responsiveness of adaptive experiments can benefit students by providing immediate feedback and quick adjustments in instruction to ensure that the student understands the material, which is particularly beneficial when dealing with complex problem-solving tasks. While some existing AI in Education technologies, such as Intelligent Tutoring Systems, focus on providing contextual feedback \cite{aleven2016help}, the adaptive experimentation approach can be useful when it is unclear which form of feedback is more efficient, which is common in practice. Here, the 'form' could be alternative explanations or even direct vs. metacognitive vs. motivational feedback \cite{ott16}.

Another prominent direction is to support students' motivation and participation, which is crucial to designing CS1 courses \cite{rafferty22}. Finding and rapidly improving ways to provide motivational support blended with feedback on student performance can help foster motivation and engagement in the classroom and promote student achievement, belonging, and self-worth.
Although theories of academic motivation in educational psychology, as well as relevant theories from social psychology, can provide good insight \cite{wigfield_where_2020} into CS classrooms, transforming them into effective support interventions may be non-trivial \cite{hagger_handbook_2020}. Adaptive experimentation can support these attempts by ensuring quick improvement based on student action feedback.

\section{Conclusions and Future Work}
\label{sec:conclusion}

In this paper, we discuss how conducting adaptive experiments over traditional ones can benefit CS1 students. We also outline promising areas for conducting adaptive experiments in CS1. Balancing instructional and research goals is a complex task, and even as instructors and researchers use adaptive experiments, many emerging needs and issues still need to be addressed at the experimental framework or process levels.

The further development of a body of research and best practices on the use of adaptive experimentation for continuous improvement in education should look at the design and action research paradigms \cite{mckenney_conducting_2018,clement_call_2004}, as well as the ideas of network improvement communities \cite{hallinan_getting_2011}, and research-practice partnerships \cite{penuel_conceptualizing_2015}. This, in turn, can contribute to answering the call for increased empirical validation in Computing Education Research~\cite{al-zubidy_updated_2016, heckman_systematic_2022}. Finally, more research should be conducted to investigate the impact of adaptive systems on long-term student outcomes, such as academic performance and retention, and to explore the potential of personalization to meet each student's unique needs.

\begin{acks}
This work was partially supported by NSF grant \#2209819, Office of Naval Research grant number \#N00014-21-1-2576, and Natural Sciences and Engineering Research Council of Canada (NSERC), (\#RGPIN-2019-06968) to Joseph Jay Williams. We are grateful to Haochen Song, Dr Stanislav Pozdniakov, and Nathan Laundry for their helpful suggestions.
\end{acks}

\balance
\bibliographystyle{ACM-Reference-Format}
\bibliography{sigcse2023-instructors-adaptive-experiments}

\end{document}